\documentclass[atmp]{ipart_v1}

\Vol{19}
\Issue{3}
\Year{2015}
\firstpage{701}

\usepackage{t1enc}
\usepackage[latin1]{inputenc}
\usepackage[english]{babel}

\usepackage{amsthm}
\usepackage{yfonts}

\usepackage{bbm}
\usepackage{bm}
\usepackage{mathrsfs}
\usepackage{array,multirow,makecell}
\usepackage{graphicx}

\newcommand{\be}[0]{\begin{equation}}
\newcommand{\ee}[0]{\end{equation}}

\numberwithin{equation}{section}

\theoremstyle{plain}% default

\begin{document}

\title[Painlev\'{e} analysis and  integrability]{Painlev\'{e} analysis and  integrability of three-dimensional Armbruster Guckenheimer Kim galactic potential}

\author[Walid Chatar et al.]{Walid Chatar$ ^{1} $, Jaouad Kharbach$ ^{1} $, Mohamed Benkhali$ ^{1} $, Mohamed Benmalek$ ^{1} $, Abdellah Rezzouk$ ^{1} $, Mohammed Ouazzani-Jamil$ ^{2} $}

\begin{abstract}
The three-dimensional Armbruster Guckenheimer Kim galactic potential in the general form is considered. The integrability of this problem is performed by using the Painlev\'{e} analysis. We report three cases of integrability, exact integrals of motion are obtained explicitly for each case. A numerical experiment is provided to investigate the classical phase space and explore chaos-order-chaos phenomenon.
\end{abstract}

\maketitle

\section{Introduction}\leavevmode\par
Dynamical systems are mathematical notions that can be used to model physical phenomena by differential equations. Using these equations, we can study the evolution over time of dynamical systems and determine their behavior: periodic, quasi-periodic or chaotic. Dynamical systems are used in many branches of physics such as in galactic dynamics.

In general, dynamical systems are described by Hamiltonian mechanics in terms of generalized coordinates $q_{i}$ and their generalized momenta $p_{i}$, these systems are known by Hamiltonian systems. In addition, most of the Hamiltonian systems are not integrable, however, completely integrable systems are very rare in nature and a small perturbation can make them not integrable or chaotic. Even if these systems are very rare, they play a very important role. They allow to test the validity of the fundamental laws of physics.

Galactic dynamics are among the most important branches of astrophysics whose development began during the last seven decades. So, most physicists and astronomers had a vision of the physical world dominated by integrable or quasi-integrable systems (Contopoulos \cite{Contopoulos2002}). Many articles have studied galactic dynamics because this area represents a great challenge for research, these articles have studied several dynamical aspects such as the integrability of a model that can describe the motion of galaxies, for exemple Armbruster \cite{Armbruster1989}, the existence of periodic orbits and their linear stability, see for instance (Llibre and Vidal \cite{Llibre and Vidal 2012}, Alfaro et al. \cite{Alfaro et al 2013}, Llibre and Makhlouf \cite{Llibre and Makhlouf 2013}, Llibre and Roberto \cite{Llibre and Roberto 2013}, Llibre et al. \cite{Llibre et al 2014}), and the regular and chaotic behaviors of the orbits, see, e.g., (Habib et al. \cite{Habib et al 1997}, Caranicolas \cite{Caranicolas2000}, Karanis and Caranicolas \cite{Karanis and Caranicolas 2001}). Moreover, chaotic phenomena could play an important role in physical systems and also for galactic dynamics, chaos has taken an important place in the studies that have been made.

Most articles study two types of motion. One of them is studied the global motion in galaxies, as in the axially symmetric mass model that was used by Caranicolas \cite{Caranicolas1996}. The other type describes the local galactic motion, and the researches in this type have been centered on the models of elliptic galaxies, because the rotation of these galaxies is made with a low angular velocity (see, e.g., Caranicolas and Barbanis \cite{Caranicolas and Barbanis 1982}, Zeeuw and Merritt \cite{Zeeuw and Merritt 1983}, Innanen \cite{Innanen 1982}).

To study the integrability of dynamical systems, that is to say the distinction between integrable and non-integrable systems, different methods and approaches exist such as the Painlev\'{e} analysis \cite{chatar2019, Elmandouh2016}, the Ziglin theorem \cite{Morales and Simo 1994}, the Liouville theorem \cite{Abraham1978}, Lie algebra \cite{Zhdanov 1998}, separability study \cite{Chatar et al. 2019b, Chatar et al. 2019c}, the Smaller Alignment index (SALI) method for chaos detection \cite{Carita et al. 2017}, and the Poincar\'{e} sections \cite{Goudas et al. 2006}.

In general, a physical system whose energy is conserved is modeled by a Hamiltonian system with $n$ degrees of freedom, this Hamiltonian system is integrable in the sense of Liouville, if there exist $n-1$ other additional integrals in involution $\lbrace H,I_{j}\rbrace=0, j=1,...,n-1 $ which are functionally independent \cite{Abraham1978}.

The purpose of this article is to study the integrability of the Hamiltonian with three degrees of freedom, given by:
\begin{align}
\label{eq:1}
H_{3D} & =\frac{1}{2}(p_{x}^{2}+p_{y}^{2}+p_{z}^{2})-\frac{\mu}{2}(x^{2}+y^{2}+z^{2})-\frac{\alpha}{4}(x^{2}+y^{2}+z^{2})^{2} \\
\notag & -\frac{\beta}{2}z^{2}(x^{2}+y^{2})
\end{align}
where $\mu,\alpha$ and $\beta$ are free arbitrary parameters, $p_{x},p_{y}$  and $p_{z}$ are generalized momenta corresponding, respectively, to the generalized coordinates $ x,y $ and $ z $.\\
The associated equations of motion are:
\begin{equation}
\label{eq:2}
      \left\{
    \begin{aligned}
    \dot{p_{x}}&=-\frac{\partial H}{\partial x}=x\left[ \mu + \alpha(x^{2}+y^{2}+z^{2})+\beta z^{2} \right], \:\: \dot{x}=\frac{\partial H}{\partial p_{x}}=p_{x}\\
      \dot{p_{y}}&=-\frac{\partial H}{\partial y}=y\left[ \mu + \alpha(x^{2}+y^{2}+z^{2})+\beta z^{2} \right], \:\: \dot{y}=\frac{\partial H}{\partial p_{y}}=p_{y}\\
      \dot{p_{z}}&=-\frac{\partial H}{\partial z}=z\left[ \mu + (\alpha +\beta) (x^{2}+y^{2})+ \alpha z^{2} \right], \: \dot{z}=\frac{\partial H}{\partial p_{z}}=p_{z}\\
    \end{aligned}
  \right.
\end{equation}
By making use cylindrical coordinates:
\begin{equation}
\label{eq:3}
\left\{
    \begin{aligned}
  x=&\rho \cos \theta, \:\: p_{x}=\cos \theta p_{\rho}-\frac{\sin \theta}{\rho} p_{\theta}\\
     y=&\rho \sin \theta, \:\: p_{y}=\sin \theta p_{\rho}+\frac{\cos \theta}{\rho} p_{\theta}\\
     z=&z, \:\: p_{z}=p_{z}\\
    \end{aligned}
  \right.
\end{equation}
The Hamiltonian of the Armbruster-Guckenheimer-Kim becomes
\begin{equation}
\label{eq:4}
 H_{3D}=\frac{1}{2}(p_{\rho}^{2}+p_{z}^{2})+\frac{k^{2}}{2\rho^{2}}-\frac{\mu}{2}(\rho^{2}+z^{2})
-\frac{\alpha}{4}(\rho^{2}+z^{2})^{2}-\frac{\beta}{2}\rho^{2}z^{2}
\end{equation}
where $p_{\rho}$  and $p_{z}$ are generalized momenta corresponding, respectively, to the generalized coordinates $\rho$ and $ z $, while $ k $ represents the value of the cyclic integral associated to the cyclic coordinate $ \theta $, where $ k=p_{\theta}=xp_{y}-yp_{x} $ represents the component of angular momentum about $ z- $axis.

For $ k=0 $, $ H_{3D} $ is equivalent to a Hamiltonian with two degrees of freedom. This Hamiltonian is known in literature as Hamiltonian of Armbruster-Guckenheimer-Kim (AGK). This potential characterizes the local motion of an almost axisymmetric galaxy that rotates with a constant angular velocity $ \omega $ around a fixed axis, for this reason (Elmandouh \cite{Elmandouh2016}) adds the term $ \omega p_{\theta} $ which can also appear in several physical problems, but most work has studied the AGK Hamiltonian for $ \omega =0 $ (as a non-rotating system), see for instance (Acosta-Humanez \cite{Acosta-Humanez 2018}, Llibre and Roberto \cite{Llibre and Roberto 2013}).

The first condition for the Hamiltonian of 2D-AGK to be integrable is $ \omega =0 $, there are three integrable cases of 2D Hamiltonian: $ \beta =2\alpha $ \cite{Elmandouh2016}, $ \beta =-\alpha $ and $ \beta =0 $ \cite{Armbruster1989}. Table \ref{tbl:1} shows the Hamiltonian expressions and the second integral of motion for each integrable case.

The article is organized as follows. In Section \ref{s:2}, we studied the integrability of the 3D-AGK problem by using the Painlev\'{e} analysis and we compared the results with the 2D-AGK problem. In section \ref{s:3}, we determined the integrables of motion. Numerical investigations are presented in Section \ref{s:4} to confirm the integrability of 3D-AGK and show the transition chaos-order-chaos of this system. Finally, this article is completed by a conclusion that summarizes the obtained results.
\begin{table}
\caption{The integrable cases and the corresponding first integrals of motion.\label{tbl:1}}
\begin{tabular*}{\textwidth}{ccc}
\hline
Case & Hamiltonian & Second invariant\\
\hline
$ \beta =0 $ & $ H_{2D}=\frac{1}{2}(p_{x}^{2}+p_{y}^{2})-\frac{\mu}{2}(x^{2}+y^{2}) $ & $ I_{2D}=(xp_{y}-yp_{x})^{2} $ \\
 & $ -\frac{\alpha}{4}(x^{2}+y^{2})^{2} $ &   \\
$ \beta =-\alpha $ & $ H_{2D}=\frac{1}{2}(p_{x}^{2}+p_{y}^{2})-\frac{\mu}{2}(x^{2}+y^{2}) $ & $ I_{2D}=p_{y}^{2}-p_{x}^{2}+\mu (x^{2}-y^{2}) $\\
& $ -\frac{\alpha}{4}(x^{4}+y^{4}) $ & $ +\frac{\alpha}{2}(x^{4}-y^{4}) $ \\
$ \beta =2\alpha $ & $ H_{2D}=\frac{1}{2}(p_{x}^{2}+p_{y}^{2})-\frac{\mu}{2}(x^{2}+y^{2}) $ & $ I_{2D}=-p_{x}p_{y}+\mu xy $\\
& $ -\frac{\alpha}{4}(x^{2}+y^{2})^{2}-\alpha x^{2}y^{2} $ &  $ +\alpha xy(x^{2}+y^{2}) $ \\
\hline
\end{tabular*}
\end{table}
\section{Painlev\'{e} analysis of 3D-AGK}\label{s:2}\leavevmode\par
The 3D-AGK is described by the following Hamiltonian:
\begin{align}
\label{eq:5}
H_{3D} & =\frac{1}{2}(p_{x}^{2}+p_{y}^{2}+p_{z}^{2})+A(x^{2}+y^{2}+z^{2})+B(x^{4}+y^{4}+z^{4})\\
\notag & +Cx^{2}y^{2}+Dz^{2}(x^{2}+y^{2})
\end{align}
where $ A=-\frac{\mu}{2}, B=-\frac{\alpha}{4}, C=-\frac{\alpha}{2}, D=-\frac{(\alpha +\beta)}{2} $\\
The associated equations of motion are
\begin{equation}
\label{eq:6}
\left\{
    \begin{aligned}
  \frac{d^{2}x}{dt^{2}}+2x\left[ A + 2Bx^{2}+Cy^{2}+Dz^{2} \right]=0 \\
      \frac{d^{2}y}{dt^{2}}+2y\left[ A + Cx^{2}+2By^{2}+Dz^{2} \right]=0 \\
      \frac{d^{2}z}{dt^{2}}+2z\left[ A + D (x^{2}+y^{2})+ 2B z^{2} \right]=0 \\
    \end{aligned}
  \right.
\end{equation}
The Painlev\'{e} analysis contains the following three steps.
\subsection{Leading-order behaviors}\leavevmode\par
Considering Eqs.(\ref{eq:6}), we assume that the leading-order behaviors of $ x(t), y(t) $ and $ z(t) $ in a sufficiently small neighborhood of the movable singularity $ t_{0} $ are 
\begin{equation}
\label{eq:7}
x(t)=a_{0}\tau^{p}, y(t)=b_{0}\tau^{q}, z(t)=c_{0}\tau^{s}, \tau =t-t_{0}\rightarrow 0
\end{equation}
and we obtain leading-order equations, by substituting (\ref{eq:7}) in Eqs.(\ref{eq:6})
\begin{equation}
\label{eq:8}
\left\{
    \begin{aligned}
  a_{0}&p(p-1)\tau^{p-2}+4Ba_{0}^{3}\tau^{3p}+2Ca_{0}b_{0}^{2}\tau^{p+2q}+2Da_{0}c_{0}^{2}\tau^{p+2s}=0\\
   b_{0}&q(q-1)\tau^{q-2}+4Bb_{0}^{3}\tau^{3q}+2Ca_{0}^{2}b_{0}\tau^{2p+q}+2Db_{0}c_{0}^{2}\tau^{q+2s}=0\\
   c_{0}&s(s-1)\tau^{s-2}+4Bc_{0}^{3}\tau^{3s}+2Dc_{0}(a_{0}^{2}\tau^{2p+s}+b_{0}^{2}\tau^{2q+s})=0\\    
    \end{aligned}
  \right.
\end{equation}
we find from Eqs.(\ref{eq:8}) the following three distinct sets of possibilities
\begin{itemize}
\item Case 1: $ p=q=s=-1 $
\end{itemize}
\begin{equation}
\label{eq:9}
\left\{
    \begin{aligned}
      2&Ba_{0}^{2}+Cb_{0}^{2}+Dc_{0}^{2}=-1\\
   2&Bb_{0}^{2}+Ca_{0}^{2}+Dc_{0}^{2}=-1\\
   2&Bc_{0}^{2}+D(a_{0}^{2}+b_{0}^{2})=-1\\
    \end{aligned}
  \right.
\end{equation}
\begin{itemize}
\item Case 2: $ p=s=-1, q=\frac{1}{2}\pm\frac{1}{2}\left[1-8(Ca_{0}^{2}+Dc_{0}^{2})\right]^{\frac{1}{2}} $
\end{itemize}
\begin{equation}
\label{eq:10}
a_{0}^{2}=c_{0}^{2}=\frac{D-2B}{4B^{2}-D^{2}},\qquad     b_{0}^{2}\mbox{: arbitrary}
\end{equation}
\begin{itemize}
\item Case 3: $ p=-1, q=\frac{1}{2}\pm\frac{1}{2}\left( 1+\frac{4C}{B}\right)^{\frac{1}{2}}, s=\frac{1}{2}\pm\frac{1}{2}\left( 1+\frac{4D}{B}\right)^{\frac{1}{2}} $
\end{itemize}
\begin{equation}
\label{eq:11}
a_{0}^{2}=-\frac{1}{2B},\qquad     b_{0}^{2} \mbox{: arbitrary,}\qquad c_{0}^{2} \mbox{: arbitrary}
\end{equation}
\subsection{Resonances}\leavevmode\par
For finding the resonances, we put this form of solutions
\begin{equation}
\label{eq:12}
x(t)=a_{0}\tau^{p}+\Omega_{1}\tau^{p+r},\quad y(t)=b_{0}\tau^{q}+\Omega_{2}\tau^{q+r},\quad z(t)=c_{0}\tau^{s}+\Omega_{3}\tau^{s+r}
\end{equation}
we substitute (\ref{eq:12}) into the equations of motion (\ref{eq:6}), and we obtain a system of linear algebraic equations: $ M_{3}(r)\Omega=0 $, $ \Omega=(\Omega_{1},\Omega_{2},\Omega_{3}) $, where $ M_{3}(r) $ is a $ 3\times3 $ matrix dependent on $ r $, we calculate $ detM_{3}(r)=0 $.\\
For case 1, $ p=q=s=-1 $
\begin{eqnarray}
\label{eq:13}
detM_{3}(r)=&
\begin{tabular}{|ccc|}
$ r'+8Ba_{0}^{2} $&$ 4Ca_{0}b_{0} $&$ 4Da_{0}c_{0} $\\ 
$ 4Ca_{0}b_{0} $&$ r'+8Bb_{0}^{2} $&$ 4Db_{0}c_{0} $\\ 
$ 4Da_{0}c_{0} $&$ 4Db_{0}c_{0} $&$ r'+8Bc_{0}^{2} $\\
\end{tabular}=0\\
r'=r^{2}-4r&\nonumber
\end{eqnarray}
It is easy to check that $ r'=4 $ is a root of (\ref{eq:13}), and so
\begin{align}
\label{eq:14}
( & r'-4)(r'+X)(r'+Y)=0 \notag \\
X & +Y=4\left[1+2B\left(a_{0}^{2}+b_{0}^{2}+c_{0}^{2} \right) \right] \\
XY=4(X+Y)+ & 16\left[(4B^{2}-C^{2})a_{0}^{2}b_{0}^{2}+(4B^{2}-D^{2})b_{0}^{2}c_{0}^{2}+4(B^{2}-D^{2})a_{0}^{2}c_{0}^{2} \right]\notag
\end{align}
From (\ref{eq:14}) we can deduce that the resonances occur at
\begin{eqnarray*}
r=-1, 4, \frac{3}{2}\pm \frac{1}{2}(9-4X)^{\frac{1}{2}}, \frac{3}{2}\pm \frac{1}{2}(9-4Y)^{\frac{1}{2}}
\end{eqnarray*}
The restriction that the resonance values (except $ -1 $) be non-negative rationals but must
depend on the nature of the leading-order singularity, leads to the following possibilities:
\begin{itemize}
\item Case 1(i)\begin{eqnarray}
\label{eq:15}
X=2,\qquad Y=2,\qquad r=-1,1,1,2,2,4
\end{eqnarray}
\end{itemize}
\begin{itemize}
\item Case 1(ii)\begin{eqnarray}
\label{eq:16}
X=2,\qquad Y=0,\qquad r=-1,0,1,2,3,4
\end{eqnarray}
\end{itemize}
\begin{itemize}
\item Case 1(iii)\begin{eqnarray}
\label{eq:17}
X=0,\qquad Y=0,\qquad r=-1,0,0,3,3,4
\end{eqnarray}
\end{itemize}
For case 2, we obtain the resonance condition after omitting the coefficients of less
dominant terms as
\begin{eqnarray*}
detM_{3}(r)=
\begin{tabular}{|ccc|}
$ r^{2}-3r+8Ba_{0}^{2} $&$ 0 $&$ 4Da_{0}c_{0} $\\ 
$ 4Ca_{0}b_{0} $&$ r(r+2q-1) $&$ 4Db_{0}c_{0} $\\ 
$ 4Da_{0}c_{0} $&$ 0 $&$ r^{2}-3r+8Bc_{0}^{2} $\\
\end{tabular}=0
\end{eqnarray*}
which on expanding gives
\begin{eqnarray*}
r=-1, 0, 4, (1-2q), \frac{1}{2}\left[3\pm (9-4Z)^{\frac{1}{2}} \right] \quad\mbox{with}\quad Z=4\left[1+2B(a_{0}^{2}+c_{0}^{2}) \right]  
\end{eqnarray*}
Further, the consideration of non-negative rational resonances in conjunction with the leading-order
behaviour leads to the following possibilities:
\begin{itemize}
\item Case 2(i)
\end{itemize}
\begin{eqnarray*}
q=0, Z=0,\quad Ca_{0}^{2}+Dc_{0}^{2}=0, B(a_{0}^{2}+c_{0}^{2})=-\frac{1}{2},\quad r=-1,0,0,1,3,4 
\end{eqnarray*}
\begin{itemize}
\item Case 2(ii)
\end{itemize}
\begin{eqnarray*}
q=0, Z=2,\quad Ca_{0}^{2}+Dc_{0}^{2}=0, B(a_{0}^{2}+c_{0}^{2})=-\frac{1}{4},\quad r=-1,0,1,1,2,4 
\end{eqnarray*}
\begin{itemize}
\item Case 2(iii)
\end{itemize}
\begin{eqnarray*}
q=-\frac{1}{2}, Z=0,\quad Ca_{0}^{2}+Dc_{0}^{2}=-\frac{3}{8}, B(a_{0}^{2}+c_{0}^{2})=-\frac{1}{2},\quad r=-1,0,0,2,3,4 
\end{eqnarray*}
\begin{itemize}
\item Case 2(iv)
\end{itemize}
\begin{eqnarray*}
q=-\frac{1}{2}, Z=2,\quad Ca_{0}^{2}+Dc_{0}^{2}=-\frac{3}{8}, B(a_{0}^{2}+c_{0}^{2})=-\frac{1}{4},\quad r=-1,0,1,2,2,4 
\end{eqnarray*}
For Case $ 3 $, in a similar way, we derive the following sets of possibilities:
\begin{itemize}
\item Case 3(i)
\end{itemize}
\begin{eqnarray*}
q=0,\quad s=0,\quad C=D=0,\quad r=-1,0,0,1,1,4 
\end{eqnarray*}
\begin{itemize}
\item Case 3(ii)
\end{itemize}
\begin{eqnarray*}
q=0,\quad s=-\frac{1}{2},\quad C=0,\quad D=\frac{3}{4}B,\quad r=-1,0,0,1,2,4 
\end{eqnarray*}
\begin{itemize}
\item Case 3(iii)
\end{itemize}
\begin{eqnarray*}
q=-\frac{1}{2},\quad s=0,\quad C=\frac{3}{4}B,\quad D=0,\quad r=-1,0,0,1,2,4 
\end{eqnarray*}
\begin{itemize}
\item Case 3(iv)
\end{itemize}
\begin{eqnarray*}
q=-\frac{1}{2},\quad s=-\frac{1}{2},\quad C=D=\frac{3}{4}B,\quad r=-1,0,0,2,2,4 
\end{eqnarray*}
\subsection{Evaluation of arbitrary constants}\leavevmode\par
Thus we have isolated eleven distinct sets of integer resonances, namely (\ref{eq:15})$ - $(\ref{eq:17}) for case 1. In order to compute the arbitrary constants for the above resonances, we introduce the series representations
\begin{eqnarray}
\label{eq:18}
x(t)=\sum_{k=0}^{4}a_{k}\tau^{p+k} , y(t)=\sum_{k=0}^{4}b_{k}\tau^{q+k}, z(t)=\sum_{k=0}^{4}c_{k}\tau^{s+k}
\end{eqnarray}
we substitute (\ref{eq:17}) into the equations of motion (\ref{eq:6}), and for each case, we obtain recursion relations for $ a_{k}, b_{k} $ and $ c_{k} $, $ k=0,1..4 $\\
For case 1, they take the following forms:
\begin{eqnarray}
&(j-1)(j-2)a_{j}+2Aa_{j-2}+4B\sum_{l}\sum_{m}a_{j-l-m}a_{l}a_{m}+2C\sum_{l}\sum_{m}b_{j-l-m}a_{l}b_{m}\nonumber\\
&+2D\sum_{l}\sum_{m}c_{j-l-m}a_{l}c_{m}=0
\label{eq:19}
\end{eqnarray}
\begin{eqnarray}
&(j-1)(j-2)b_{j}+2Ab_{j-2}+4B\sum_{l}\sum_{m}b_{j-l-m}b_{l}b_{m}+2C\sum_{l}\sum_{m}a_{j-l-m}b_{l}a_{m}\nonumber\\
&+2D\sum_{l}\sum_{m}c_{j-l-m}b_{l}c_{m}=0
\label{eq:20}
\end{eqnarray}
\begin{eqnarray}
&(j-1)(j-2)c_{j}+2Ac_{j-2}+4B\sum_{l}\sum_{m}c_{j-l-m}c_{l}c_{m}+2D\sum_{l}\sum_{m}a_{j-l-m}c_{l}a_{m}\nonumber\\
&+2D\sum_{l}\sum_{m}b_{j-l-m}c_{l}b_{m}=0
\label{eq:21}
\end{eqnarray}
where $ 0\leq l,m \leq j \leq 4 $, by solving (\ref{eq:19})$ - $(\ref{eq:21}), one can obtain the various $ a_{k}, b_{k} $ and $ c_{k} $ explicitly.\\
For example, for $ j = 0 $
\begin{equation*}
\left\{
    \begin{aligned}
2&Ba_{0}^{2}+Cb_{0}^{2}+Dc_{0}^{2}+1=0\\
   2&Bb_{0}^{2}+Ca_{0}^{2}+Dc_{0}^{2}+1=0\\
   2&Bc_{0}^{2}+D(a_{0}^{2}+b_{0}^{2})+1=0\\
    \end{aligned}
  \right.
\end{equation*}
For $ j = 1 $
\begin{equation*}
\left\{
    \begin{aligned}
6&Ba_{0}^{2}a_{1}+C(2a_{0}b_{0}b_{1}+a_{1}b_{0}^{2})+D(2a_{0}c_{0}c_{1}+a_{1}c_{0}^{2})=0\\
   6&Bb_{0}^{2}b_{1}+C(2a_{0}a_{1}b_{0}+a_{0}^{2}b_{1})+D(2b_{0}c_{0}c_{1}+b_{1}c_{0}^{2})=0\\
   6&Bc_{0}^{2}c_{1}+D(a_{0}^{2}c_{1}+b_{0}^{2}c_{1}+2a_{0}a_{1}c_{0}+2b_{0}b_{1}c_{0})=0\\
    \end{aligned}
  \right.
\end{equation*}
and similarly for $ j=2,3,4 $, we determine the relations between the constants of the system such as the coefficients $ a_{k}, b_{k} $ or $ c_{k} $ become arbitrary and we respect the conditions indicated by the resonance values in (\ref{eq:15})$ - $(\ref{eq:17}). Then, we get the following parametric restrictions:
\begin{itemize}
\item Case 1(i): $ C=2B,\quad D=0 $,\quad $ \forall (A, B) $\quad $ \Leftrightarrow\quad \beta =-\alpha $,\quad $ \forall (\mu ,\alpha ) $
\end{itemize}
\begin{itemize}
\item Case 1(ii): $ C=2B,\quad D=6B $,\quad $ \forall (A, B) $\quad $ \Leftrightarrow\quad \beta =2\alpha $,\quad $ \forall (\mu ,\alpha ) $
\end{itemize}
\begin{itemize}
\item Case 1(iii): $ C=D=2B $,\quad $ \forall (A, B) $\quad $ \Leftrightarrow\quad \beta =0 $,\quad $ \forall (\mu ,\alpha ) $
\end{itemize}
Proceeding analogously as before for case 2 we find: 
\begin{equation*}
\left\{
    \begin{aligned}
B&=16B\\
    C&=D=12B\\
    D&=2B\\
    A&=4A\\
    \end{aligned}
  \right.
  \quad\mbox{and}\quad
    \left\{
    \begin{aligned}
B&=8B\\
    C&=D=6B\\
    D&=2B\\
    A&=4A\\
    \end{aligned}
  \right.
\end{equation*}
For case 3 we find:
\begin{equation*}
\left\{
    \begin{aligned}
B&=16B\\
    C&=D=12B\\
    D&=6B\\
    A&=4A\\
    \end{aligned}
  \right.
  \quad\mbox{and}\quad
    \left\{
    \begin{aligned}
B&=8B\\
    C&=0\\
    D&=0\\
    A&=4A\\
    \end{aligned}
  \right.
\end{equation*}
Then, these conditions are not logical, and hence, this cases 2 and 3 are omitted.
\section{The integrals of motion of 3D-AGK}\label{s:3}\leavevmode\par
In this section, we investigate the explicit form of the second and third integrals of motion, in order to substantiate the investigations of the Painlev\'{e} analysis of the section \ref{s:2}.  For this purpose, let us consider the spherical coordinates $ (r,\theta ,\varphi) $, the Hamiltonian (\ref{eq:1}) becomes
\begin{equation}
\label{eq:22}
 H_{3D}=\frac{1}{2}(p_{r}^{2}+\frac{p_{\theta}^{2}}{r^{2}})+\frac{f^{2}}{2r^{2}\sin ^{2}\theta}-\frac{\mu}{2}r^{2}-\frac{\alpha}{4}r^{4}-\frac{\beta}{2}r^{4}\cos ^{2}\theta \sin ^{2}\theta
\end{equation}
where $ p_{r} $ and $ p_{\theta} $ are generalized momenta corresponding, respectively, to the generalized coordinates $ r $ and $ \theta $, while $ f $ represents the value of the cyclic integral associated to the cyclic coordinate $ \varphi $
\begin{equation}
\label{eq:23}
I_{1}=p_{\varphi}=f
\end{equation}
Moreover, we can determine the cyclic variable $ \varphi $ from $ \varphi -\varphi_{0}=\int\frac{\delta H}{\delta f}dt $.\\
The Hamilton equations associated with the Hamiltonian (\ref{eq:22}) are given by
\begin{equation}
\label{eq:24}
 \left\{
    \begin{aligned}
    \dot{p_{r}}&=\frac{p_{\theta}^{2}}{r^{3}}+\frac{f^{2}}{r^{3}\sin ^{2}\theta}+\mu r+r^{3}(\alpha +2\beta \cos ^{2}\theta \sin^{2}\theta ), \quad \dot{r}=p_{r}\\
      \dot{p_{\theta}}&=\frac{f^{2}\cos \theta}{r^{2}\sin ^{3}\theta}+\beta r \cos\theta\sin\theta (\cos ^{2}\theta -\sin ^{2}\theta ), \quad \dot{\theta}=\frac{p_{\theta}}{r^{2}}\\
    \end{aligned}
  \right.
\end{equation}
The integral of energy corresponds to the equations Hamilton (\ref{eq:24})
\begin{equation}
\label{eq:25}
H_{3D}=\frac{1}{2}(p_{r}^{2}+\frac{p_{\theta}^{2}}{r^{2}})+\frac{f^{2}}{2r^{2}\sin ^{2}\theta}-\frac{\mu}{2}r^{2}-\frac{\alpha}{4}r^{4}-\frac{\beta}{2}r^{4}\cos ^{2}\theta \sin ^{2}\theta =h
\end{equation}
where $ h $ is a constant representing the value of the energy integral.

According to Jacobi's theory, the Hamilton equations (\ref{eq:24}) are completely integrable, if there exists an integral of the motion $ I_{3D} $ independent of the integral of the energy. This integral is known in literatures under the complementary or additional integral name. In general, it is well known that complementary integrals are either quadratic or quartic in terms of momenta. For 3D-AGK, if the complementary integral is quadratic, it takes the form (\ref{eq:26}), but if it is quartic, it takes the form (\ref{eq:27}).
\begin{equation}
\label{eq:26}
I_{3D}=I_{2D}+f(\Lambda_{3}p_{r}^{2}+\Lambda_{2}p_{r}p_{\theta}+\Lambda_{1}p_{\theta}^{2}+\Lambda_{0})
\end{equation}
\begin{equation}
\label{eq:27}
I_{3D}=I_{2D}^{2}+f(\Lambda_{3}p_{r}^{2}+\Lambda_{2}p_{r}p_{\theta}+\Lambda_{1}p_{\theta}^{2}+\Lambda_{0})
\end{equation}
Where $ I_{2D} $ is the complementary integral of 2D-AGK indicated in Table \ref{tbl:1} (Using spherical coordinates $ (r,\theta ,\varphi) $) and $ \Lambda_{i}(i=0...3) $ being functions in the generalized coordinates $ r,\theta $. Notice that the order of the complementary integral (\ref{eq:27}) is doubled and that we can name doubled order. This situation is found in several problems (see, e.g., Wojciechowski \cite{Wojciechowski1985}, Grammaticos \cite{Grammaticos1985}). Note that the quartic integral (\ref{eq:27}) becomes quadratic when $ f=0 $.\\
Using the derivative of $ I_{3D} $ with respect to time $ t $ or using the Poisson bracket $\lbrace H_{3D},I_{3D}\rbrace $, by matching the coefficients of moments to zero, using Hamilton equations (\ref{eq:24}), we obtain a system of nonlinear partial differential equations whose solution gives the expressions of $ \Lambda_{i}(r,\theta) $. Thus, the complementary integrals of the three cases are
\begin{itemize}
\item Case 1(i) : $ \beta =- \alpha $
\end{itemize}
The complementary integral $ I_{3D}$ is quadratic in terms of the momenta
\begin{align}
\label{eq:28}
I_{3D}=I_{2D}+f\left[ p_{r}^{2}+\frac{1}{r^{2}}p_{\theta}^{2}+\frac{f(f-1)}{r^{2}\sin ^{2}\theta}-\mu r^{2} - \frac{\alpha}{2}r^{4}(1-2\cos ^{2}\theta \sin ^{2}\theta)\right]
\end{align}
\begin{itemize}
\item Case 1(ii) : $ \beta =2 \alpha $
\end{itemize}
The complementary integral $ I_{3D}$ is quartic in terms of the momenta
\begin{align}
\label{eq:29}
I_{3D}=I_{2D}^{2}+f\left( \frac{f\cos ^{2}\theta}{r^{2}\sin ^{2}\theta}p_{r}^{2} -\frac{2f\cos \theta}{r^{3}\sin \theta}p_{r}p_{\theta}+\frac{f}{r^{4}}p_{\theta}^{2}-2f\alpha  r^{2} \cos ^{2}\theta \right)
\end{align}
\begin{itemize}
\item Case 1(iii) : $ \beta =0 $
\end{itemize}
The complementary integral $ I_{3D}$ is quartic in terms of the momenta
\begin{equation}
\label{eq:30}
I_{3D}=(p_{\theta}^{2}+\frac{f^{2}}{\sin ^{2}\theta})^{2}
\end{equation}
It is clear that on a zero level of the cyclic integral $ f=p_{\varphi}=0 $, the complementary integral of each case is reduced to the quadratic integral of 2D-AGK for the same case. In addition, the formula of complementary integrals (\ref{eq:28}), (\ref{eq:29}) and (\ref{eq:30}) are not unique, because any function $ Q(I_{1},H_{3D}) $ can be added to these integrals without its accuracy being affected, because $ \dot{Q}(I_{1},H_{3D})=0 $
\section{Numerical illustration}\label{s:4}\leavevmode\par
Using a set of software routines, to plot 3D trajectories of motion in the phase space for each case, we give numerical illustrations to confirm the three-dimensional integrability of the system. The trajectories of motion are plotted in the phase space $ (p_{y},y,p_{x}) $ for cases 1(i) and 1(iii), in the phase space $ (p_{x},z,y) $ for case 1(ii), we can also choose another phase space to plot the trajectories of motion.

\begin{figure}
\begin{center}
\begin{tabular}{@{}ccc@{}}
\includegraphics[scale=0.4]{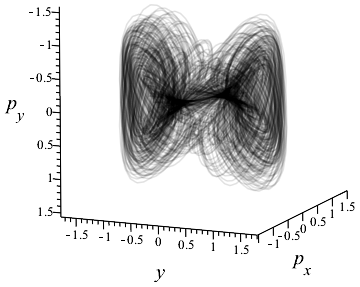}   & \includegraphics[scale=0.4]{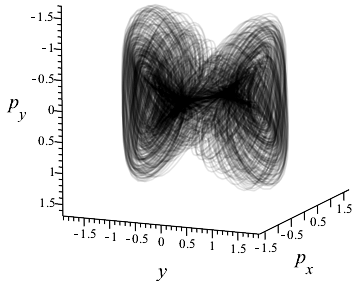} & \includegraphics[scale=0.4]{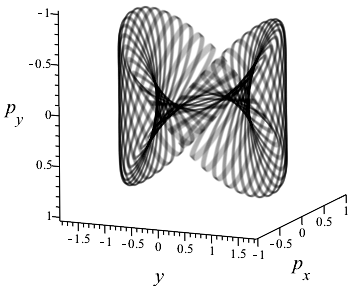}\\
\begin{footnotesize}a\end{footnotesize} & \begin{footnotesize}b\end{footnotesize} & \begin{footnotesize}c\end{footnotesize}
\end{tabular}
\begin{tabular}{@{}cc@{}}
\includegraphics[scale=0.4]{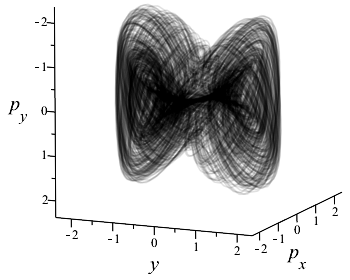}   & \includegraphics[scale=0.4]{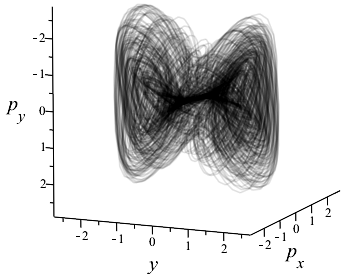} \\
\begin{footnotesize}d\end{footnotesize} & \begin{footnotesize}e\end{footnotesize}
\end{tabular}
\caption{Trajectories of motion in the phase space for case 1(i): $ \mu=3 $, $ \beta=2 $ and (a)  $ \alpha=-2.4, h=-0.7219 $; (b) $ \alpha=-2.2, h=-0.7429 $; (c) $ \alpha=-2, h=-0.7639 $; (d) $ \alpha=-1.8, h=-0.7849 $; (e) $ \alpha=-1.6, h=-0.8059 $}
\end{center}
\label{fig:1}
\end{figure}
\begin{figure}[th]
\begin{center}
\begin{tabular}{@{}ccc@{}}
\includegraphics[scale=0.4]{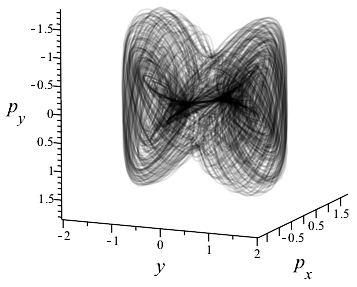}   & \includegraphics[scale=0.4]{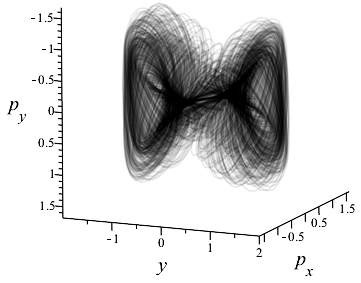} & \includegraphics[scale=0.4]{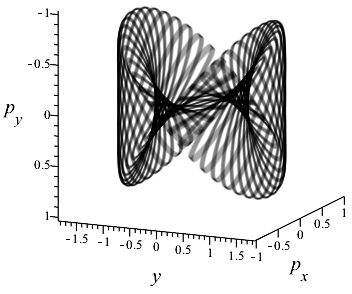}\\
\begin{footnotesize}a\end{footnotesize} & \begin{footnotesize}b\end{footnotesize} & \begin{footnotesize}c\end{footnotesize}
\end{tabular}
\begin{tabular}{@{}cc@{}}
\includegraphics[scale=0.4]{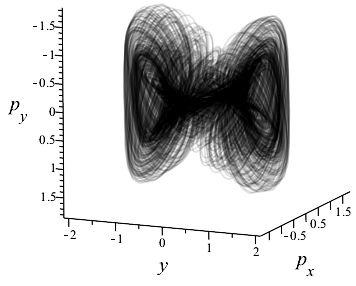}   & \includegraphics[scale=0.4]{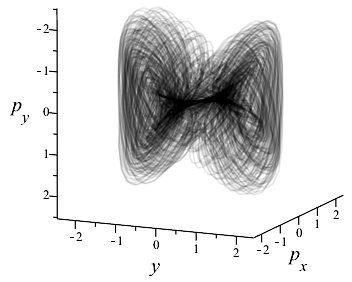} \\
\begin{footnotesize}d\end{footnotesize} & \begin{footnotesize}e\end{footnotesize}
\end{tabular}
\caption{Trajectories of motion in the phase space for case 1(i): $ \mu=3 $, $ \alpha=-2 $ and (a) $ \beta=1.75, h=-0.7633 $; (b) $ \beta=1.9, h=-0.7636 $; (c) $ \beta=2, h=-0.7639 $; (d) $ \beta=2.1, h=-0.7641 $; (e) $ \beta=2.5, h=-0.7652 $}
\end{center}
\label{fig:2}
\end{figure}
\begin{figure}[th]
\begin{center}
\begin{tabular}{@{}ccc@{}}
\includegraphics[scale=0.4]{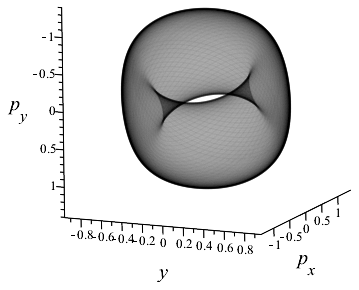}   & \includegraphics[scale=0.4]{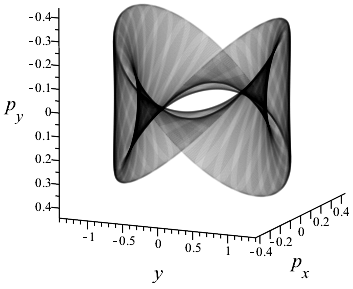} & \includegraphics[scale=0.4]{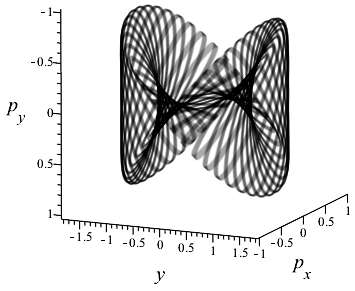}\\
\begin{footnotesize}a\end{footnotesize} & \begin{footnotesize}b\end{footnotesize} & \begin{footnotesize}c\end{footnotesize}
\end{tabular}
\begin{tabular}{@{}cc@{}}
\includegraphics[scale=0.4]{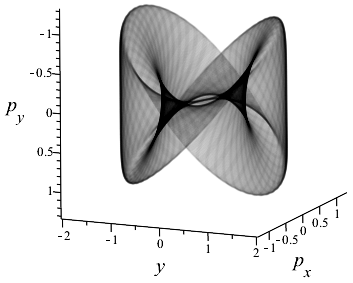}   & \includegraphics[scale=0.4]{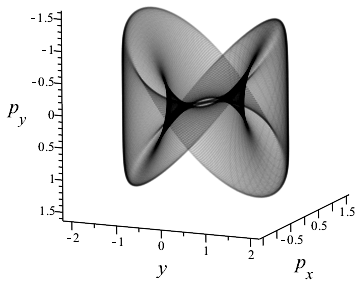} \\
\begin{footnotesize}d\end{footnotesize} & \begin{footnotesize}e\end{footnotesize}
\end{tabular}
\caption{Trajectories of motion in the phase space for case 1(i): $ \alpha=-2 $, $ \beta=2 $ and (a) $ \mu=-1.5, h=0.6939 $; (b) $ \mu=2, h=-0.4399 $; (c) $ \mu=3, h=-0.7639 $; (d) $ \mu=3.5, h=-0.9259 $; (e) $ \mu=4, h=-1.0879 $}
\end{center}
\label{fig:3}
\end{figure}
Figures 1, 2 and 3 illustrate the trajectories of motion around case 1(i), we kept two parameters and we varied the third to observe the chaos-order-chaos transition.

In Figure 1, we fixed $ \mu=3 $, $ \beta=2 $ and we varied $ \alpha $, for $ \alpha=-2 $ the integrability of the system is satisfied, as it is shown in Figure 1(c), the trajectories of motion are completely regular, if we let us move a little away from the integrable case that is to say for $ \alpha=-2.2 $ and $ \alpha=-1.8 $, the system becomes mixed: ordered and chaotic simultaneously, which is clearly visible in Figures 1(b,d), these two values of $ \alpha $ present the beginning of chaos when the value of $ \alpha $ is varied in both directions $ \alpha <-2 $ and $ \alpha >-2 $, when we further increase the value of $ \alpha $ in both directions $ \alpha=-2.4 $ and $ \alpha=-1.6 $, we observe that the chaotic regions become more dominant than the regular regions which is clearly visible in Figures 1(a,e). In Figure 2, we kept the parameters $ \mu=3 $, $ \alpha=-2 $ and we varied the parameter $ \beta $ in both directions $ \beta <2 $ and $ \beta >2 $ to observe the chaos-order-chaos transition as it is shown in Figures 2(a-e), since the chaos in the system is most strongly observed in the vicinity of $ \beta=1.75 $ and $ \beta=2.5 $. In the same way, in Figure 3, we kept the parameters $ \alpha=-2 $, $ \beta=2 $ and we varied the parameter $ \mu $, as it is shown in Figures 3(a-e), the chaos-order-chaos phenomenon is not observed because the system is integrable $ \forall\mu $ in case 1(i), i.e. the trajectories of motion are completely regular.

\begin{figure}[th]
\begin{center}
\begin{tabular}{@{}ccc@{}}
\includegraphics[scale=0.4]{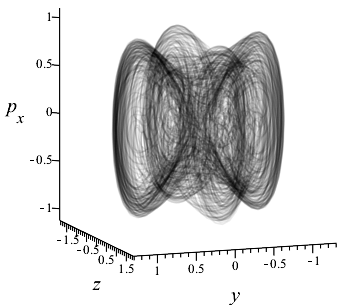}   & \includegraphics[scale=0.4]{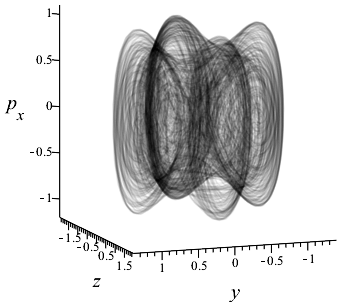} & \includegraphics[scale=0.4]{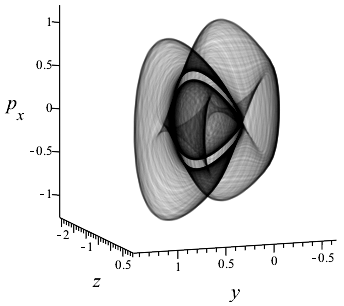}\\
\begin{footnotesize}a\end{footnotesize} & \begin{footnotesize}b\end{footnotesize} & \begin{footnotesize}c\end{footnotesize}
\end{tabular}
\begin{tabular}{@{}cc@{}}
\includegraphics[scale=0.4]{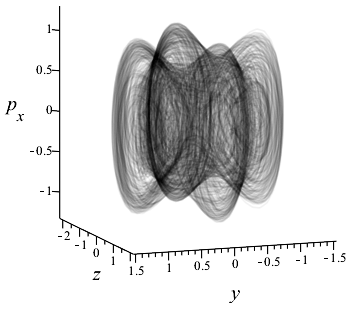}   & \includegraphics[scale=0.4]{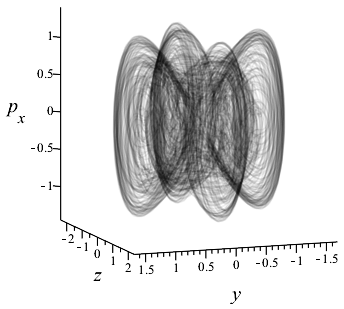} \\
\begin{footnotesize}d\end{footnotesize} & \begin{footnotesize}e\end{footnotesize}
\end{tabular}
\caption{Trajectories of motion in the phase space for case 1(ii): $ \mu=3 $, $ \beta=-4 $ and (a)  $ \alpha=-2.5, h=-0.2297 $; (b) $ \alpha=-2.2, h=-0.2306 $; (c) $ \alpha=-2, h=-0.2311 $; (d) $ \alpha=-1.8, h=-0.2317 $; (e) $ \alpha=-1.5, h=-0.2326 $}
\end{center}
\label{fig:4}
\end{figure}
\begin{figure}[th]
\begin{center}
\begin{tabular}{@{}ccc@{}}
\includegraphics[scale=0.4]{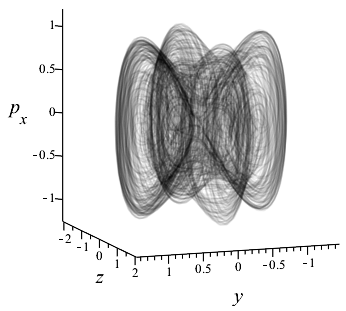}   & \includegraphics[scale=0.4]{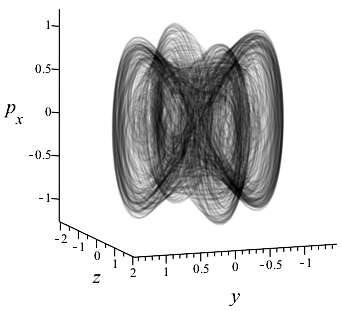} & \includegraphics[scale=0.4]{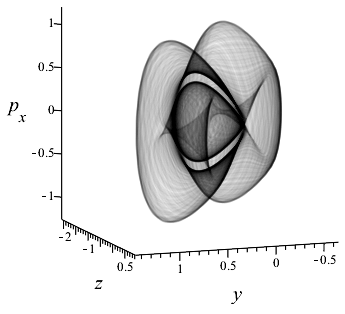}\\
\begin{footnotesize}a\end{footnotesize} & \begin{footnotesize}b\end{footnotesize} & \begin{footnotesize}c\end{footnotesize}
\end{tabular}
\begin{tabular}{@{}cc@{}}
\includegraphics[scale=0.4]{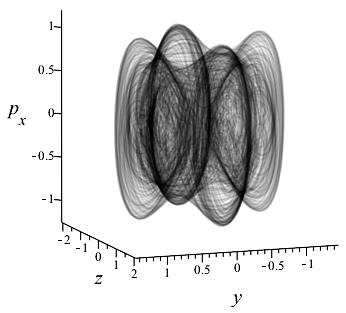}   & \includegraphics[scale=0.4]{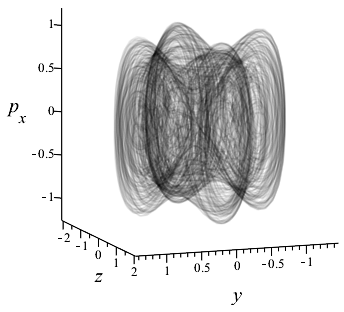} \\
\begin{footnotesize}d\end{footnotesize} & \begin{footnotesize}e\end{footnotesize}
\end{tabular}
\caption{Trajectories of motion in the phase space for case 1(ii): $ \mu=3 $, $ \alpha=-2 $ and (a) $ \beta=-4.5, h=-0.23115 $; (b) $ \beta=-4.3, h=-0.23117 $; (c) $ \beta=-4, h=-0.23119 $; (d) $ \beta=-3.8, h=-0.23120 $; (e) $ \beta=-3.5, h=-0.23122 $}
\end{center}
\label{fig:5}
\end{figure}
\begin{figure}[th]
\begin{center}
\begin{tabular}{@{}ccc@{}}
\includegraphics[scale=0.4]{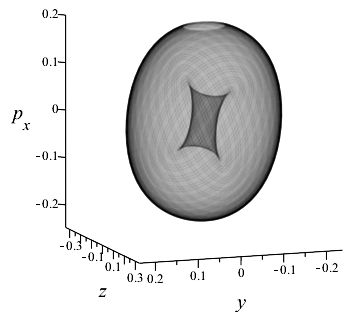}   & \includegraphics[scale=0.4]{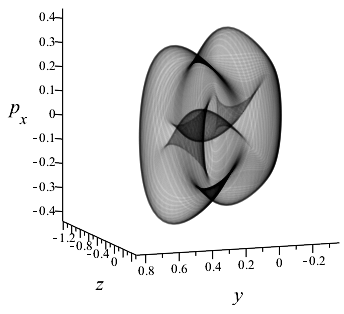} & \includegraphics[scale=0.4]{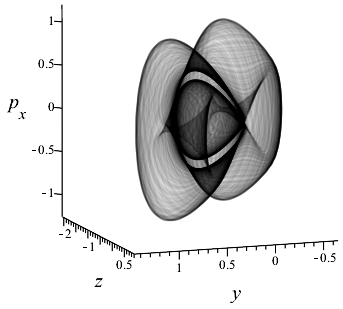}\\
\begin{footnotesize}a\end{footnotesize} & \begin{footnotesize}b\end{footnotesize} & \begin{footnotesize}c\end{footnotesize}
\end{tabular}
\begin{tabular}{@{}cc@{}}
\includegraphics[scale=0.4]{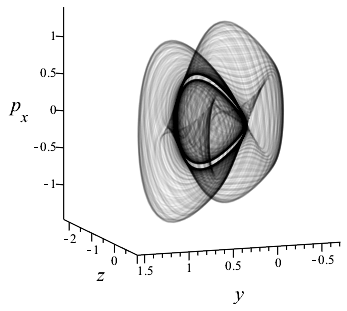}   & \includegraphics[scale=0.4]{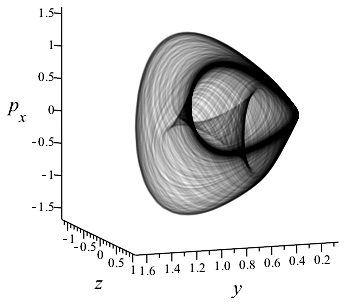} \\
\begin{footnotesize}d\end{footnotesize} & \begin{footnotesize}e\end{footnotesize}
\end{tabular}
\caption{Trajectories of motion in the phase space for case 1(ii): $ \alpha=-2 $, $ \beta=-4 $ and (a) $ \mu=-1, h=0.4528 $; (b) $ \mu=1, h=0.1108 $; (c) $ \mu=3, h=-0.2311 $; (d) $ \mu=3.5, h=-0.3166 $; (e) $ \mu=4, h=-0.4021 $}
\end{center}
\label{fig:6}
\end{figure}
Around the case 1(ii), the Figures 4, 5 and 6 illustrate the trajectories of motion for different values of the parameters of the system $ \alpha$, $\beta $ and $ \mu $.

In Figure 4, we fixed the parameters $ \mu=3$, $\beta=-4$ and we varied the parameter $ \alpha $ to know the influence of this parameter on the system, for $ \alpha=-2 $ the system is integrable as it is shown in Figure 4(c), the trajectories of motion are very regular. For $ \alpha=-2.2 $ and $ \alpha=-1.8 $ as it is shown in Figures 4(b,d), we observe the beginning of chaos in both directions $ \alpha <-2 $ and $ \alpha >-2 $, that is to say, chaotic regions are observed beside regular regions. When the value of $ \alpha $ is increased in both directions $ \alpha=-2.5 $ and $ \alpha=-1.5 $ as it is shown in Figures 4(a,e), we observe that the chaotic regions which have become more dominant than the regular regions. In Figure 5, we kept the parameters $ \mu=3$, $\alpha=-2 $ and varying the parameter $ \beta $, we observe that the chaos-order-chaos phenomenon is distinctly visible in the trajectories of motion, the system is integrable for $ \beta=-4 $ as it is shown in Figure 5(c), the system is in the beginning of chaos for $ \beta=-4.3 $ and $ \beta=-3.8 $ as it is shown in Figures 5(b,d), we observe more chaotic regions for $ \beta=-4.5 $ and $ \beta=-3.5 $ as it is shown in Figures 5(a,e). If we vary the parameter $ \mu $, the system is always integrable as it is shown in Figures 6(a-e).

\begin{figure}[th]
\begin{center}
\begin{tabular}{@{}ccc@{}}
\includegraphics[scale=0.4]{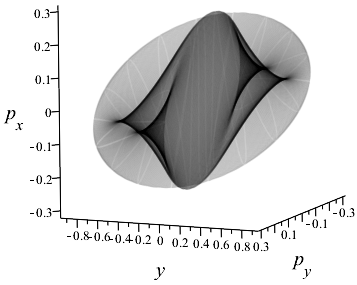}   & \includegraphics[scale=0.4]{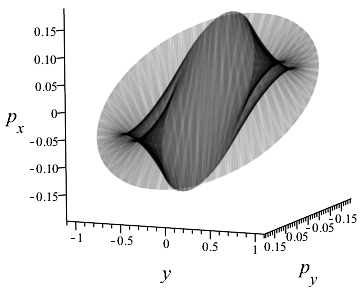} & \includegraphics[scale=0.4]{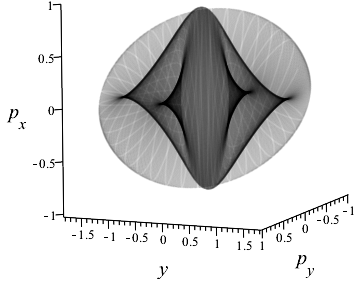}\\
\begin{footnotesize}a\end{footnotesize} & \begin{footnotesize}b\end{footnotesize} & \begin{footnotesize}c\end{footnotesize}
\end{tabular}
\begin{tabular}{@{}cc@{}}
\includegraphics[scale=0.4]{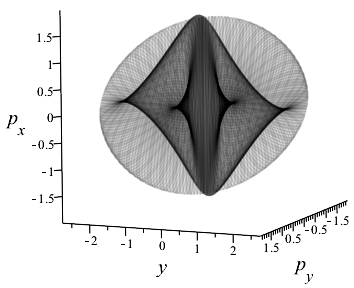}   & \includegraphics[scale=0.4]{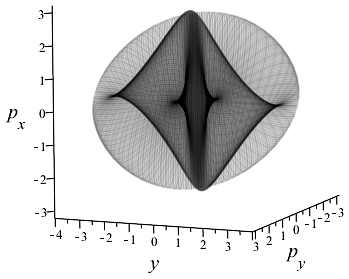} \\
\begin{footnotesize}d\end{footnotesize} & \begin{footnotesize}e\end{footnotesize}
\end{tabular}
\caption{Trajectories of motion in the phase space for case 1(iii): $ \mu=3 $, $ \beta=0 $ and (a)  $ \alpha=-6, h=-0.3390 $; (b) $ \alpha=-4, h=-0.5489 $; (c) $ \alpha=-2, h=-0.7588 $; (d) $ \alpha=-1, h=-0.8638 $; (e) $ \alpha=-0.5, h=-0.9162 $}
\end{center}
\label{fig:7}
\end{figure}
\begin{figure}[th]
\begin{center}
\begin{tabular}{@{}ccc@{}}
\includegraphics[scale=0.4]{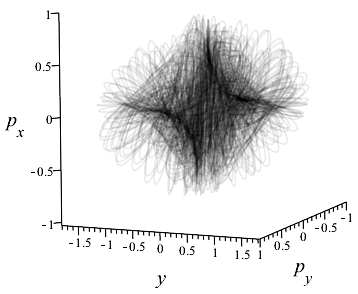}   & \includegraphics[scale=0.4]{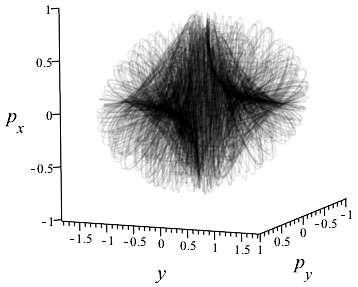} & \includegraphics[scale=0.4]{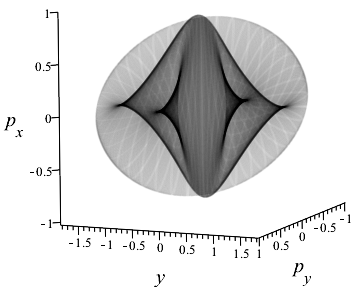}\\
\begin{footnotesize}a\end{footnotesize} & \begin{footnotesize}b\end{footnotesize} & \begin{footnotesize}c\end{footnotesize}
\end{tabular}
\begin{tabular}{@{}cc@{}}
\includegraphics[scale=0.4]{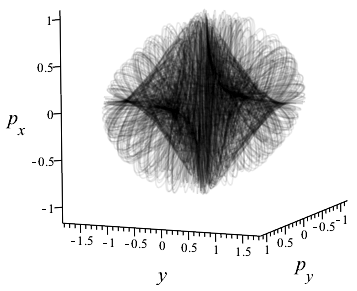}   & \includegraphics[scale=0.4]{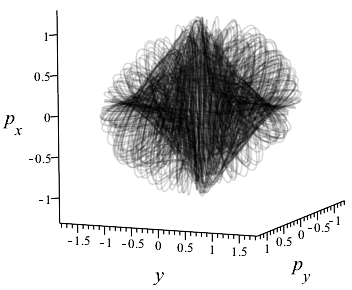} \\
\begin{footnotesize}d\end{footnotesize} & \begin{footnotesize}e\end{footnotesize}
\end{tabular}
\caption{Trajectories of motion in the phase space for case 1(iii): $ \mu=3 $, $ \alpha=-2 $ and (a) $ \beta=-1, h=-0.7563 $; (b) $ \beta=-0.5, h=-0.7575 $; (c) $ \beta=0, h=-0.7588 $, (d) $ \beta=0.4, h=-0.7598 $; (e) $ \beta=0.7, h=-0.7606 $}
\end{center}
\label{fig:8}
\end{figure}
\begin{figure}[th]
\begin{center}
\begin{tabular}{@{}ccc@{}}
\includegraphics[scale=0.4]{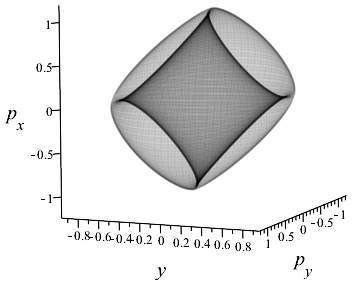}   & \includegraphics[scale=0.4]{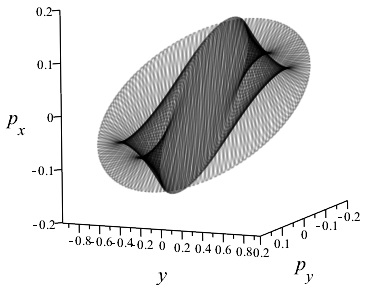} & \includegraphics[scale=0.4]{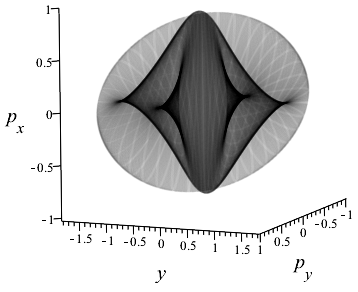}\\
\begin{footnotesize}a\end{footnotesize} & \begin{footnotesize}b\end{footnotesize} & \begin{footnotesize}c\end{footnotesize}
\end{tabular}
\begin{tabular}{@{}cc@{}}
\includegraphics[scale=0.4]{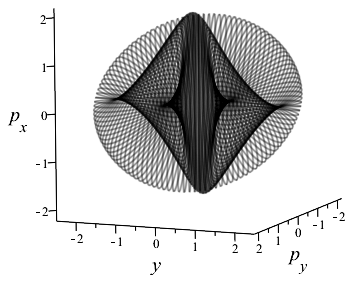}   & \includegraphics[scale=0.4]{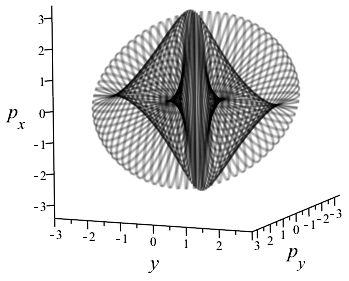} \\
\begin{footnotesize}d\end{footnotesize} & \begin{footnotesize}e\end{footnotesize}
\end{tabular}
\caption{Trajectories of motion in the phase space for case 1(iii): $ \alpha=-2 $, $ \beta=0 $ and (a) $ \mu=-1, h=0.5370 $; (b) $ \mu=1, h=-0.1109 $; (c) $ \mu=3, h=-0.7588 $; (d) $ \mu=5, h=-1.4068 $; (e) $ \mu=7, h=-2.0547 $}
\end{center}
\label{fig:9}
\end{figure}
In the same way as the two cases above, we kept two parameters and we varied the third, to show the influence of each parameter on the integrability of the system. Figures 7, 8 and 9 illustrate the trajectories of motion around case 1(iii) for different values of the parameters of the system $ \alpha$, $\beta $ and $ \mu $.

The system is always integrable when we varied the parameters $ \alpha$ and $ \mu $ because the condition of integrability in case 1(iii) is $\beta=0 $, that is to say does not depend on the parameters $ \alpha$ and $ \mu $, so the chaos-order-chaos phenomenon is not observed. To confirm this result we varied these parameters, as it is shown in Figures 7(a-e) for $ \alpha$ and Figures 9(a-e) for $ \mu $. The chaos-order-chaos transition is shown in Figure 8 when we varied the parameter $ \beta $. For $\beta=0 $ as it is shown in Figure 8(c), the integrability of the system is satisfied, for $\beta=-0.5 $ and $\beta=0.4 $ as it is shown in Figures 8(b,d) we have the beginning of chaos, and we observe chaotic regions in the vicinity of regular regions, the chaotic behavior increases when $\beta=-1 $ and $\beta=0.7 $ as it is shown in Figures 8(a,e).

According to the Figures 1-9 that represent the trajectories of motion around the three integrable cases of the system, when the parameters $ \alpha $, $ \beta $ and $ \mu $ are varied, we show that chaos-order-chaos transition is a robust phenomenon, i.e. the mechanism of the transition changes rapidly if one of the parameters of the system is varied, except for the parameter $ \mu $ and the parameter $ \alpha $ for case 1 (iii). We deduce that the parameter $ \mu $ does not change the integrability of the system for the three cases. For the variation of the parameter $ \alpha $, the behavior of the system is changed in the vicinity of the integrable cases except for case 1(iii), the variation of parameter $ \beta $ shows the chaos-order-chaos transition for the three integrable cases of the system, i.e. a small perturbation allows to change the behavior of system.
\section{Conclusion}\leavevmode\par
To conclude, we have studied the integrability of 3-dimensional Armbruster-Guckenheimer-Kim Hamiltonian 3D-AGK, for this reason we determined the integrable cases by using the Painlev\'{e} analysis and we found three cases as in the 2D-AGK. In order to confirm integrability  of the 3D-AGK, we have accompanied this analysis  by the construction of the additional first integrals of motion for each case. Only in this way, the powerful predictive nature of the method will be materialized. However, the possibility of existence of other integrable cases far from the known ones should not be ruled out until the contrary is proven.

Finally, we have explored the chaos-order-chaos phenomenon  in detail by working with the variation of the system control parameters. We deduce that the parameter $ \mu $ does not change the integrability of the system for the three cases, but the system is very sensitive to the variation of the parameter $ \beta $ for the three cases and admits a strong sensitivity for the parameters $ \alpha $ in cases 1(i) and 1(ii), but the variation of $ \alpha $ does not influence the system in case 1(iii).

\providecommand{\href}[2]{#2}

\address{
$ ^{1} $Laboratoire de Physique du Solide, Facult\'{e} des Sciences Dhar El Mahraz, Universit\'{e} Sidi Mohamed Ben Abdellah, B.P. 1796, 30000 Fez-Atlas, Morocco.\\
$ ^{2} $Laboratoire Syst\`{e}mes et Environnements Durables, Universit\'{e} Priv\'{e}e de F\`{e}s, Lot. Quaraouiyine Route Ain Chkef, 30040 Fez, Morocco.\\
\email{walid.chatar@usmba.ac.ma}\\
}
\end{document}